\documentclass[journal,twoside,web]{ieeecolor}

\usepackage{lcsys}

\usepackage{cite}
\usepackage{amsmath,amssymb,amsfonts,mathrsfs}
\usepackage{algorithmic}
\usepackage{graphicx}
\usepackage{algorithm,algorithmic}
\usepackage{hyperref}
\usepackage{ulem}   
\hypersetup{hidelinks=true}
\usepackage{textcomp}
\usepackage{relsize}
\usepackage{caption}

\newcommand\copyrighttext{%
  \footnotesize © 2025 IEEE. Personal use of this material is permitted. Permission from IEEE must be obtained for all other uses, in any current or future media, including reprinting/republishing this material for advertising or promotional purposes, creating new collective works, for resale or redistribution to servers or lists, or reuse of any copyrighted component of this work in other works.}
\newcommand\copyrightnotice{%
\begin{tikzpicture}[remember picture,overlay]
\node[anchor=north,yshift=-10pt] at (current page.north) {\fbox{\parbox{\dimexpr\textwidth-\fboxsep-\fboxrule\relax}{\copyrighttext}}};
\end{tikzpicture}%
}

\usepackage{tikz,pgfplots}
\usetikzlibrary{arrows.meta, shapes.geometric, shadows, positioning, calc, backgrounds}

\usepgfplotslibrary{patchplots}
\usepgfplotslibrary{fillbetween}
\pgfplotsset{%
     layers/standard/.define layer set={%
         background,axis background,axis grid,axis ticks,axis lines,axis tick labels,pre main,main,axis descriptions,axis foreground%
     }{
         grid style={/pgfplots/on layer=axis grid},%
         tick style={/pgfplots/on layer=axis ticks},%
         axis line style={/pgfplots/on layer=axis lines},%
         label style={/pgfplots/on layer=axis descriptions},%
         title style={/pgfplots/on layer=axis descriptions},%
         colorbar style={/pgfplots/on layer=axis descriptions},%
         ticklabel style={/pgfplots/on layer=axis tick labels},%
         axis background@ style={/pgfplots/on layer=axis background},%
         3d box foreground style={/pgfplots/on layer=axis foreground},%
     },
 }

\newtheorem{lemma}{Lemma}

\newtheorem{proposition}{Proposition}
\newtheorem{assumption}{Assumption}
\newtheorem{corollary}{Corollary}
\newtheorem{theorem}{Theorem}
\newtheorem{remark}{Remark}
\newtheorem{example}{Example}
\newtheorem{definition}{Definition}

\newcommand{\changes}[1]{{\color{black}#1}}

\def\BibTeX{{\rm B\kern-.05em{\sc i\kern-.025em b}\kern-.08em
    T\kern-.1667em\lower.7ex\hbox{E}\kern-.125emX}}
\begin{document}
\title{Nash Equilibrium Learning In Large Populations With First-Order Payoff Modifications}
\author{Matthew S. Hankins, Jair Cert\'{o}rio, Tzuyu Jeng, Nuno C. Martins
\thanks{This work was supported by 
AFOSR Grant FA9550-23-1-0467 and 
NSF Grants 2139713 and  
2135561.}%
\thanks{Matthew S. Hankins, Jair Cert\'orio, Tzuyu Jeng, Nuno C. Martins are with the Dept. of ECE and the ISR at the University of Maryland, College Park (e-mail: \{msh,certorio,tyjeng,nmartins\}@umd.edu). }%
}

\maketitle
\copyrightnotice
\thispagestyle{empty}

\begin{abstract}
\changes{We establish Nash equilibrium learning in large populations of noncooperative, strategic agents. Our analysis considers the broadest class to date of payoff mechanisms with first-order modifications, capable of modeling bounded rationality and anticipatory effects, averaging, or Pad\'{e} delay approximations. We propose a framework that, for the first time, combines two nonstandard system-theoretic passivity notions. Our results hold for discontinuous best response dynamics alongside continuous learning rules, significantly extending prior work.}
\end{abstract}



\section{Introduction}
\label{sec:introduction}

\IEEEPARstart{P}{opulation} games and evolutionary dynamics provide a tractable framework for analyzing large populations of learning agents \cite{Sandholm2010Population-Game,Sandholm2015Handbook-of-gam}. Learning rules model how agents revise their strategies based on the current population strategy profile and the payoffs associated with each available strategy. \changes{Agents may be natural or synthetic, as seen in~\cite{Sandholm2015Handbook-of-gam,Quijano2017The-role-of-pop}.}

First-order payoff modifications are an important tool in modeling prevalent learning behaviors and information dissemination effects \cite{Arslan2006Anticipatory-le,Fox2013Population-Game,Park2018Payoff-Dynamic-,Gao2020On-Passivity-Re,Mabrok2021Passivity-Analy}.
However, prior work lacks a general treatment of first-order payoff modifications and has been limited to replicator dynamics~\cite{Mabrok2021Passivity-Analy} or specific classes of payoff mechanisms, such as those based on strictly concave potential games~\cite[\S9.B]{Park2018Payoff-Dynamic-} and contractive affine games~\cite[\S4.3.3]{Fox2013Population-Game}.

\changes{Recent studies have examined Nash equilibrium learning~\cite{Martins_2025aa,Park2018Payoff-Dynamic-} via nonstandard passivity concepts, including counterclockwise dissipativity~\cite{Angeli2006Systems-With-Co} and $\delta$-passivity~\cite{Fox2013Population-Game}. This paper is the first to apply these two notions jointly, establishing convergence results for the broadest class of payoff mechanisms with first-order modifications and evolutionary dynamics considered thus far. We consider payoff dynamics models belonging to a generalization of potential games based on counterclockwise passivity and we allow evolutionary learning dynamics resulting from combinations of continuous learning rules alongside the discontinuous best response dynamic. \\}

\noindent {\bf Notation:} \changes{ For $A\in\mathbb{C}^{n\times m}$, we use $A'$ to denote the transpose conjugate of $A$, while $A \succeq 0$, and  $A \succ 0$ denote, respectively, that $A$ is positive semidefinite or positive definite.}

\section{Framework and Evolutionary Dynamics}
\label{sec:EDIM}

Our framework models the noncooperative strategic interactions of many agents. For simplicity, we assume the agents belong to a single population characterized by a set of available strategies $\{1,\ldots,n\}$.
Each agent adopts one strategy at a time, which the agent may revise repeatedly. The payoff vector $p(t)$ in $\mathbb{R}^n$, whose $i$-th entry $p_i(t)$ quantifies the net rewards of the strategy $i$ at time $t$, influences the revision process, as the agents typically seek strategies with higher payoffs. The agents are nondescript and the population state $x(t)$ in 
${\mathbb{X}:= \Big \{x \in [0,1]^n \ \big | \ \textstyle\sum_{\ell=1}^n x_\ell =1 \Big \}}$
approximates the population's aggregate strategic choices in the large population limit~\cite{Sandholm2003Evolution-and-e}. Namely, $x_i(t)$ approximates the proportion of the population's agents following strategy $i$ at time $t$. 

\subsection{Learning Rules And Evolutionary Dynamics}

\changes{We consider population state trajectories $\mathbf{x}$ and payoff vector trajectories $\mathbf{p}$ that are Lipschitz continuous with respect to time, leading to sets of trajectories $\mathscr{X}$ and $\mathscr{P}$ defined as
\begin{align*}
\mathscr{X}:=&\Big \{ \mathbf{x}:[0,\infty)\to \mathbb{X}  \ \ \big | \  \text{$\mathbf{x}$ { is Lipschitz continuous} } \Big \},\\
\mathscr{P}:=&\Big \{ \mathbf{p}:[0,\infty)\to \mathbb{R}^n  \ \big | \  \text{$\mathbf{p}$ { is Lipschitz continuous}} \ \Big \}.
\end{align*} }
The revision process causes $\mathbf{x}$ to vary over time. Before describing in \S\ref{subsec:evolutionary_differential_inclusion} the evolutionary differential inclusion that specifies how this process takes place, we proceed to define two types of learning behavior models and the associated maps that will determine the evolution of $\mathbf{x}$.

\subsubsection{Best response rule and associated evolutionary dynamics}

We will define the best response rule via the set-valued best response map \(\mathfrak{B}: \mathbb{R}^n \to 2^{\mathbb{X}}\), given by
\begin{equation} \label{eq:bestresponsemap}
\mathfrak{B}(p) := \arg \max_{x \in \mathbb{X}} x'p, \quad p \in \mathbb{R}^n.
\end{equation}

For \(\mathbf{p} \in \mathscr{P}\), a population exclusively following the best response rule evolves according to a state trajectory $\mathbf{x}$ in $\mathscr{X}$ satisfying the differential inclusion 
\begin{equation} \tag{BR}
\dot{x}(t) \in \mathcal{B}(x(t),p(t))
\end{equation}
for almost all \(t \geq 0\), where \(\mathcal{B}(x,p) := \{ \hat{x} - x \mid \hat{x} \in \mathfrak{B}(p) \}\). Specifically, (BR) can be expressed for almost all $t\geq 0$ as
\begin{equation} \tag{BR'}
\dot{x}(t) = \hat{x}(t) - x(t), \text{ for some $\hat{x}(t)$ in $\mathfrak{B}\big(p(t)\big)$}.
\end{equation}

 According to (BR), the term \( (\hat{x}_i(t)- x_i(t)) \) represents the net rate of agents switching to strategy \(i\), which can be interpreted as an influx ($\hat{x}_i(t)$) and outflow (\( x_i(t)\)) of agents to optimal strategies. If \(\hat{x}_{i^*}(t) > 0\) then the strategy \(i^*\) is ``optimal" at time $t$, that is, \(p_{i^*}(t) \geq p_\ell(t)\) for all $\ell$ in $\{ 1, \ldots, n\}$. When a strategy \(j\) is not optimal, then \(\hat{x}_j(t) = 0\) and strategy $j$ is abandoned according to \(\dot{x}_j(t) = - x_j(t)\).

\changes{As specified, (BR) was first studied in~\cite{Gilboa1991Social-Stabilit,Matsui_1992aa} and further discussed with relevant insights in~\cite[\S 13.5.2.2]{Sandholm2015Handbook-of-gam}.}

 The following proposition indicates that solutions to (BR) have a tendency to progress towards increasing payoffs.

\begin{proposition} \label{prop:PCNSish} For all $x$ in $\mathbb{X}$ and $p$ in $\mathbb{R}^n$, it holds that 
\begin{equation}\label{eq:PCForBR} p'\mathcal{B}(\changes{x,p}) \geq 0 \text{ and } \left[ p'\mathcal{B}(\changes{x,p}) = 0 \Leftrightarrow x \in \mathfrak{B}(p) \right].
\end{equation}
\end{proposition}

 \vspace{.02in}

\textbf{Proof:} We proceed by modifying  the proof of~\cite[Theorem~6.1.4]{Sandholm2010Population-Game}. Since $p'\mathfrak{B}(p) = \max_{1 \leq i \leq n} p_i=:\bar{p}$, we have $p'\mathcal{B}(x,p) = \bar{p}-p'x$. The proof follows from the facts that $\bar{p} \geq x'p$ for all $x$ in $\mathbb{X}$ and $x'p = \bar{p} \Leftrightarrow x \in \mathfrak{B}(p)$.~$\square$

\subsubsection{Continuous rules and associated evolutionary dynamics}
A different type of learning behavior is modeled as a continuous evolutionary dynamic~\cite[\S4.1.2]{Sandholm2010Population-Game} according to which, for a given $\mathbf{p}$ in $\mathscr{P}$, the state trajectory $\mathbf{x}$ is the solution in $\mathscr{X}$ of
\begin{equation} \tag{EDM}
\dot{x}(t) = \mathcal{V}(x(t),p(t)), \quad t \geq 0,
\end{equation} where $\mathcal{V}:\mathbb{X} \times \mathbb{R}^n \rightarrow \mathbb{R}^n$ is specified for each strategy $i$ as
\begin{equation} \label{eq:eqforV} \underbrace{\mathcal{V}_i ( x,p )}_{\text{\small net flow into $i$}} := \sum_{j=1}^n \underbrace{x_j \mathcal{T}_{ji}(x,p)}_{\text{\small flow} \ j\rightarrow i} - \underbrace{x_i \mathcal{T}_{ij}(x,p)}_{\text{\small flow} \ i\rightarrow j}.
\end{equation} Here, $\mathcal{V}_i$ is the net flow of agents switching to strategy $i$ according to a locally Lipschitz continuous map ${\mathcal{T}: \mathbb{X} \times \mathbb{R}^{n} \to \mathbb{R}_{\geq 0}^{n \times n}}$, referred to as {\it learning rule}\footnote{\changes{Learning rules are also known as strategy-revision protocols.}} (\uline{rule for short}) that models the agents' strategy revision preferences. In \cite[Part~II]{Sandholm2010Population-Game} and \cite[\S13.3-13.5]{Sandholm2015Handbook-of-gam} there is a comprehensive discussion on rule types and the classes of bounded rationality behaviors they model. In our work we will focus on a large class of rules satisfying the following assumption.

\begin{assumption}{\bf [Well-behaved rules]} \label{ass:WellBehaved} We assume that $\mathcal{T}$ is such that, for all $x$ in $\mathbb{X}$ and $p$ in $\mathbb{R}^n$, $\mathcal{V}$ in (\ref{eq:eqforV}) satisfies
\begin{align} \label{eq:PC} \tag{PC} \mathcal{V}(x,p) \neq 0 & \implies 
 p'\mathcal{V}(x,p) > 0  \\ \label{eq:NS}\tag{NS}
  \mathcal{V}(x,p)=0 &\Longleftrightarrow x \in \mathfrak{B}(p),
\end{align} where PC denotes positive correlation~\cite[\S13.4.2]{Sandholm2015Handbook-of-gam}, meaning that $\mathcal{T}$ tends to increase the population’s average payoff, and NS denotes Nash stationarity,\footnote{\changes{See Remark~\ref{rem:NSExplanation} for the origin of the term.}} meaning equilibria occur at $(x,p)$ where $x$ is a best response to $p$. We call $\mathcal{T}$ {\bf well-behaved}~\cite{Sandholm2005Excess-payoff-d} if $\mathcal{V}$ satisfies both (PC) and (NS).
\end{assumption}

As we will see in \S\ref{sec:main_results}, well-behaved rules play a key role, as (PC) and (NS) together, under stability conditions, enable a systematic characterization of the population state's limit set using $\mathfrak{B}$. Without them, this characterization may not be possible. \changes{In \S\ref{sec:ExamplesOfWellBehaved} we will discuss several important examples of well-behaved rules relevant for this paper.}

\changes{The passivity-based approach in \cite{Mabrok2021Passivity-Analy} considers \changes{first-order} payoff modifications but focuses on the replicator rule, which does not satisfy (NS).} Thus, characterizing the limit set via $\mathfrak{B}$, as done here, is not applicable in that setting.

\subsection{Evolutionary differential inclusion}
\label{subsec:evolutionary_differential_inclusion}

In our analysis henceforth, we will consider the following generalization of (BR) and (EDM) in which agents are allowed to behave in a hybrid way that combines best response as well as a continuous learning rule. 
\begin{definition} \changes{ Given $\alpha \geq 0$ and $\beta \geq 0$, with ${\alpha+\beta>0}$, the evolutionary differential inclusion model (EDIM) is}
\begin{align}
\label{eq:EDMDiffInclusion} \tag{EDIM}
\dot{x}(t) &\in \mathfrak{V}\big( x(t),p(t) \big), \quad t \geq 0, \\ \label{eq:EDIMb}
\mathfrak{V}\big( x(t),p(t) \big) &: = \alpha \mathcal{B}\big( x(t),p(t) \big) + \beta \mathcal{V}\big( x(t),p(t) \big), 
\end{align} 
where $\mathcal{T}$ is a well-behaved rule when $\beta>0$. \changes{In (\ref{eq:EDIMb}), for simplicity, we adopt an abuse of notation} to denote ${\mathfrak{V}(x,p):=\{ \alpha \tilde{x} + \beta \mathcal{V}(x,p) \ | \ \tilde{x} \in \mathcal{B}(x,p) \}}$. For a given $\mathbf{p}$ in $\mathscr{P}$, the state trajectory $\mathbf{x}$ in $\mathscr{X}$ is a solution satisfying (EDIM) for almost all $t \geq 0$.
\end{definition}
\subsubsection{Correlation function and properties}
Given $\mathfrak{V}$, we define the so-called correlation function $\rho: \mathbb{X} \times \mathbb{R}^n \rightarrow \mathbb{R}$ as
\begin{equation}
\rho(x,p):= p'\mathfrak{V}( x,p ), \ x \in \mathbb{X}, \ p \in \mathbb{R}^n.
\end{equation}

The following proposition states important properties of~$\rho$. 

\begin{proposition} \label{prop:rho} Given $\mathfrak{V}$ as in (EDIM), the following hold:
\begin{itemize}
    \item[a)] The correlation function $\rho: \mathbb{X} \times \mathbb{R}^n \rightarrow \mathbb{R}$ is a single-valued locally Lipschitz continuous map.\footnote{A generalization for constrained best response is in proof of~\cite[Thm~1]{Chen_2024aa}.}
    \item[b)] For all $x$ in $\mathbb{X}$ and $p$ in $\mathbb{R}^n$ it holds that \begin{equation} \label{eq:NSForRho}
    \rho(x,p) \geq 0 \text{ and } [\rho(x,p)=0 \Leftrightarrow x \in \mathfrak{B}(p)].
    \end{equation}
\end{itemize}
\end{proposition}
\vspace{.05in}
\textbf{Proof:} To prove a), notice that $p'\mathfrak{B}(p) = \max_{1 \leq i \leq n} p_i=:\bar{p}$, so $\rho(x,p)=p'\mathfrak{V}(x,p) = \alpha(\bar{p}-p'x)+\beta p'\mathcal{V}(x,p)$, which is a sum of locally Lipschitz continuous functions. Given that $\mathcal{T}$ is well-behaved, b) follows \changes{as (\ref{eq:PC}) requires that $p'\mathcal{V}(x,p)$ is always nonnegative and $0$ if and only if $\mathcal{V}(x,p)=0$, which holds exactly when $x\in\mathfrak{B}(p)$ by (\ref{eq:NS}). With (\ref{eq:PCForBR}), the same holds for $p'\mathcal{B}(x,p)$, and, thus, $\rho(x,p)$.}~$\square$

From (EDIM) and (\ref{eq:NSForRho}), we conclude that unless $x(t)$ is already a best response to $p(t)$, it holds that $\dot{x}(t)'p(t) > 0$, which indicates that $\dot{x}(t)$ forms an acute angle with $p(t)$, that is, $x(t)$ progresses in the direction of increasing payoffs.

\subsection{Well-behaved rule classes and Examples}
\label{sec:ExamplesOfWellBehaved}
\changes{
\begin{example} \label{example:Smith} Smith’s rule~\cite{Smith1984The-stability-o} is given by ${\mathcal{T}_{ij}^{\mathrm{\tiny Smith}}(x,p) := [p_j - p_i]_+}$, where $[\tau]_+ := \max \{ 0, \tau \}$. The corresponding $\mathcal{V}^{\mathrm{\tiny Smith}}$, from~(\ref{eq:eqforV}), has been used in traffic assignment studies~\cite{Beckmann1956Studies-in-the-}. It models agents switching from strategy $i$ to $j$ based on positive payoff differences.
\end{example}

More generally, any  {\bf impartial pairwise comparison} rule, as defined below, is well-behaved \cite[Theorem~1]{Sandholm2010Pairwise-compar}.
\begin{definition} A rule $\mathcal{T}$ is of the impartial pairwise comparison class {\bf (IPC)}~\cite[\S4.1 and \S6]{Sandholm2010Pairwise-compar} if, for all $i$ and $j$ in $\{1,\ldots,n\}$, it can be written as ${\mathcal{T}_{ij}(x,p) \underset{\tiny IPC}{:=}  \psi_j(p_j-p_i)}$. Here, $\psi_j(\tau)>0$ when $\tau>0$ and $\psi_j(\tau)=0$, otherwise. Let $\mathcal{T}$ be of the class nondecreasing IPC (or { \bf ND-IPC} for short)~\cite{Certorio_2024aa} if each $\psi_j$ is nondecreasing. 
\end{definition}

\begin{example} \label{ex:BNN} The Brown–von Neumann–Nash (BNN) rule~\cite{Brown1950Solutions-of-ga} is defined as ${\mathcal{T}_{ij}^{\mathrm{\tiny BNN}}(x,p) := [\hat{p}_j]_+}$, where $\hat{p}_i := p_i - p'x$ is the excess payoff. Unlike Smith’s rule, BNN models more conservative agents who switch only to strategies with above-average payoffs.
\end{example}

The BNN is a particular case of the so-called {\bf separable excess payoff target} rule class, as defined below, which consists of well-behaved rules \cite[Theorem 3.1]{Sandholm2005Excess-payoff-d}.

\begin{definition} A rule $\mathcal{T}$ is of the separable excess payoff target class {\bf (SEPT)}~\cite[\S2.2 and (3)]{Sandholm2005Excess-payoff-d} if for all $i$ and $j$ in $\{1,\ldots,n\}$, it can be written as ${\mathcal{T}_{ij}(x,p) \underset{\tiny SEPT}{:=}  \phi_j(\hat{p}_j)}$. Here, $\phi_j(\tau)>0$ when $\tau>0$ and $\phi_j(\tau)=0$, otherwise.
\end{definition}

The classes of IPC, ND-IPC and SEPT rules are themselves convex cones of well-behaved rules and~\cite[Theorem~2]{Sandholm2010Pairwise-compar} shows that conic combinations of rules belonging to different classes also yield well-behaved rules, often referred to as hybrid rules. Hence, well-behaved hybrid rules can model a wide range of behaviors (see also discussion in~\cite[\S5]{Sandholm2010Pairwise-compar}).

\begin{example} \label{ex:hybrid} The following rule is an example of a {\bf hybrid rule} that is well-behaved according to~\cite[Theorem~2]{Sandholm2010Pairwise-compar}:
\begin{equation*}
\mathcal{T}^{\tiny (1)}_{ij}(x,p)=\underbrace{0.2[p_j-p_i]_+ + 0.1[p_j-p_i]_+^2}_{\mathrm{ND-IPC}} + \underbrace{0.4[\hat{p}_j]_+^3 }_{\mathrm{SEPT}}
\end{equation*}
\end{example} }

\section{Payoff Mechanism And Nash Equilibrium} \label{sec:payoffmech}

 A \underline{payoff mechanism} $\mathfrak{P}:\mathscr{X} \to \mathscr{P}$ is a causal map generating an output $\mathbf{p}$ in $\mathscr{P}$ for each input $\mathbf{x}$ in $\mathscr{X}$. The roles for $\mathbf{x}$ and $\mathbf{p}$ are reversed for an (EDIM), where $\mathbf{x}$ is the output and $\mathbf{p}$ is the input.  

 Given $\mathbf{x} \in \mathscr{X}$, the following systems specify $\mathfrak{P}:\mathbf{x} \mapsto \mathbf{p}$
\begin{align} \dot{q}(t) & = \mathcal{G}_\mathfrak{F}\big( q(t),x(t) \big), \ q(0)=0, \ t \geq 0, \tag{PDMa}\\ \tag{PDMb}
u(t) & = \mathcal{H}_\mathfrak{F} \big( q(t),x(t) \big), \\ \tag{FOPMa}
\dot{s}(t)& = u(t) - \mu s(t), \ s(0)=0, \\ \tag{FOPMb}
p(t) &= \gamma \big( s(t) + \nu u(t) \big),
\end{align} where $\mathcal{G}_\mathfrak{F}:\mathbb{R}^\chi \times \mathbb{X} \rightarrow \mathbb{R}^\chi$ and $\mathcal{H}_\mathfrak{F}:\mathbb{R}^\chi \times \mathbb{X} \rightarrow \mathbb{X}$ are Lipschitz continuous maps and $\chi$ is the dimension of $\mathbf{q}$. The payoff dynamics model (PDM) above has been used in~\cite{Fox2013Population-Game,Park2019From-Population}, but here is followed by a first-order payoff modification (FOPM), with $\mu>0$ \changes{and static gain $\gamma(\nu+\mu^{-1})\ge0$}.

We view (PDM) acting  as a map $\mathfrak{F}:\mathbf{x} \mapsto \mathbf{u}$ that models how payoffs are assigned to strategies and (FOPM) specifies a map $\mathfrak{S}:\mathbf{u} \mapsto \mathbf{p}$ that could be used to model the dynamics of information dissemination, such as smoothing or first-order Pad\'{e} delay approximation, or it could also be viewed as higher-order learning behavior of the agents. Hence, the payoff mechanism is the composition $\mathfrak{P}:=\mathfrak{S} \circ \mathfrak{F}$. First-order payoff modifications were proposed for memoryless $\mathfrak{F}$ in~\cite{Fox2013Population-Game} and~\cite{Park2018Payoff-Dynamic-}, for affine $\mathfrak{F}$ and nonlinear potential games subject to additional technical conditions, respectively.

\begin{figure}
\vspace{0.04in}
\begin{center}
\scalebox{0.9}{
\begin{tikzpicture}[
                scale = 0.8,
                transform shape,
                node distance=0.5cm,
                block/.style={rectangle, draw, rounded corners, minimum width=2.3cm, minimum height=1.2cm, align=center},
                arrow/.style={->, >=stealth, ultra thick}
                ]

                \node[block, fill=teal!8] (strat) at (0,0) {\large $\dot{x} \in \mathfrak{V}(x,p)$\\{\large (EDIM)}};
                \node[block, fill=orange!8] (pdm) at (1.7,-1.5) {\large $\mathfrak{F}:\mathbf{x} \mapsto \mathbf{u}$ \\ {\large (PDM)}};
                \node[block, fill=blue!6] (fopm) at (-1.7,-1.5) {\large $\mathfrak{S}:\mathbf{u} \mapsto \mathbf{p}$ \\ {\large (FOPM)}};
                
                \draw[-{Latex[length=3mm, width=2mm]}, thick] (fopm.west) -- ++(-0.5,0) |- (strat.west) node[midway, left] {\Large $\mathbf{p}$};
                \draw[-{Latex[length=3mm, width=2mm]}, thick] (strat.east) -- ++(2.5,0) |- (pdm.east) node[midway, right] {\Large $\mathbf{x}$};
                \draw[-{Latex[length=3mm, width=2mm]}, thick] (pdm.west) -- (fopm.east) node[midway, above] {\Large $\mathbf{u}$};
                
\end{tikzpicture}
}

\end{center}
\caption{(EDIM) in feedback with payoff mechanism $\mathfrak{P}=\mathfrak{S}\circ\mathfrak{F}$.}
\label{fig:closedloop}
\end{figure}

\subsection{Solutions and a Key Assumption}
 The state trajectory of the  system formed by connecting (PDM), (FOPM) and (EDIM) in feedback (see~Figure~\ref{fig:closedloop}) is $(\mathbf{q},\mathbf{s},\mathbf{x})$, which is a Carath\'{e}odory solution for a given $x(0)$ in $\mathbb{X}$. Notice that, according to~\cite[Proposition~S2]{Cortes_2008aa}, such a solution is guaranteed to exist\changes{, though it may not be unique,} because $\mathfrak{V}$, $\mathcal{G}_\mathfrak{F}$, and $\mathcal{H}_\mathfrak{F}$ are locally bounded, and $\mathfrak{V}$ takes values in a non-empty, compact, and convex set.

\changes{By} Rademacher's Theorem~\changes{\cite[\S7.7]{Wheeden1977Measure-}}, \changes{all} $\mathbf{x}$ in $\mathscr{X}$ and $\mathbf{p}$ in $\mathscr{P}$ are differentiable for almost all time $t \geq 0$. We will be implicitly using this fact when writing integrals with respect to time of integrands involving functions of $\dot{\mathbf{x}}$ and $\dot{\mathbf{p}}$. For this reason, henceforth, all integrals are in the sense of Lebesgue. 

\begin{assumption}
\label{ass:PayoffMech} We assume that any payoff mechanism $\mathfrak{P}$ considered henceforth satisfies the following conditions.
\begin{enumerate}
    \item There is a Lipschitz continuous map $\mathcal{F}_\mathfrak{P}:\mathbb{X} \to \mathbb{R}^n$ such that for every $\mathbf{x}$ in $\mathscr{X}$, the following holds:
    $$ \lim_{t \rightarrow \infty} \| \dot{x}(t) \|=0 \implies \lim_{t \rightarrow \infty} \big \| p(t)-\mathcal{F}_{\mathfrak{P}} \big(x(t)\big) \big \| = 0.$$ We refer to $\mathcal{F}_{\mathfrak{P}}$ as the \uline{stationary game} of $\mathfrak{P}$.
    \item For the map $\mathfrak{F}:\mathbf{x} \mapsto \mathbf{u}$ specified by (PDM) there is $\beta_{\mathfrak{F}} > 0$ such that
    $$ \sup_{t \geq 0} \| u(t) \|_\infty \leq \beta_{\mathfrak{F}}, \ \mathfrak{F}:\mathbf{x} \mapsto \mathbf{u}, \ \mathbf{x} \in \mathscr{X}.$$
\end{enumerate}
\end{assumption}

Assumption~\ref{ass:PayoffMech}.1 is akin to the existence of an input-to-state characteristic~\cite{Angeli2006Systems-With-Co}. Since $\|\mathbf{x}\|_{\infty} \leq 1$, Assumption~\ref{ass:PayoffMech}.2 is satisfied for any bounded-input bounded-output (BIBO) stable $\mathfrak{F}$. Notice that because (FOPM) is also BIBO stable, Assumption~\ref{ass:PayoffMech}.2 implies that there is $\beta_{\mathfrak{P}} > 0$ such that $ \| p(t) \|_\infty <  \beta_{\mathfrak{P}}$ for all $t \geq 0$.

\subsection{Potential and Contractive Games}\label{subsubsec:pot_contr} We call $\mathfrak{F}$ a {\it game} when for all $t\geq 0$ it acts as a memoryless map ${\mathcal{F}:x(t) \mapsto p(t)}$, where $\mathcal{F}:\mathbb{X} \to \mathbb{R}^n$ is Lipschitz continuous. Here, stability and the static gain of (FOPM) imply that $ \mathcal{F}_\mathfrak{P}(x)= \gamma(\nu+\mu^{-1})\mathcal{F}(x)$ for all $x$ in $\mathbb{X}$.
\begin{definition} \label{def:game}  We refer to a game $\mathcal{F}$ as a \uline{potential game} when there is a potential function $f:\mathbb{X} \to \mathbb{R}$ satisfying\footnote{See~\cite{Sandholm2001Potential-games} and references therein for a detailed discussion.}
\begin{equation}
\label{eq:PotentialGame}
f\big ( x(T) \big) - f\big ( x(0) \big) = \int_0^T \dot{x}'(t)\mathcal{F}\big ( x(t) \big) dt,  \ \mathbf{x} \in \mathscr{X}.
\end{equation}
\end{definition} 
 \vspace{.01 in}

 Hence, $\mathcal{F}(x)=Ax+b$ is a potential game for any $b$ in $\mathbb{R}^n$ and $A=A'$ in $\mathbb{R}^{n \times n}$, with $f(x) = \tfrac{1}{2} x'Ax + b'x$. 

\begin{definition} \label{def:contr}
    We refer to a game $\mathcal{F}$ as \uline{contractive} when
    $$ (y-x)'(\mathcal{F}(y)-\mathcal{F}(x))\le 0,\quad x,y\in\mathbb{X}. $$
\end{definition}

Any potential game with a concave potential function is a contractive game \cite{Hofbauer2007Stable-games}. The congestion game in~\cite[Example~13.2]{Sandholm2015Handbook-of-gam} is a well-known example of a nonlinear contractive potential game. \changes{However, a contractive game need not be a potential game, such as symmetric zero-sum games~\cite[Example 3.3.3]{Sandholm2010Population-Game}.}

\subsection{Linear Time Invariant Payoff Mechanism}
The payoff mechanism could be linear time invariant, as defined below.

\begin{definition} \label{def:LTIPM} A payoff mechanism $\mathfrak{P}$ is of the linear time invariant {\bf (LTI)} type if $\mathfrak{F}$, specified by (PDM), is LTI. A PDM $\mathfrak{F}$ is LTI if it has a proper rational $n \times n$ transfer function matrix $F(s)$ whose poles have negative real part so as to satisfy Assumption~\ref{ass:PayoffMech}. The associated stationary game is ${\mathcal{F}_\mathfrak{P}(x)=\gamma(\nu+\mu^{-1})F(0)x}$, for all $x$ in $\mathbb{X}$.
\end{definition} 


\subsection{Nash Equilibria Set And Best Response Map} Using $\mathcal{F}_\mathfrak{P}$ \changes{as} specified in Assumption~\ref{ass:PayoffMech} for a given $\mathfrak{P}$, we can define the associated Nash equilibrium set as follows.

\begin{definition}
Given a payoff mechanism $\mathfrak{P}$, we define the {\bf Nash equilibria} set\cite{Jr.1951Non-Cooperative} of \changes{the stationary game} $\mathcal{F}_\mathfrak{P}$ as:
\begin{equation*}
\mathbb{NE}(\mathcal{F}_\mathfrak{P}) : = \Big \{ x \in \mathbb{X} \ \big | \ x \in \mathfrak{B} \big (\mathcal{F}_\mathfrak{P}(x) \big ) \Big \},
\end{equation*}
where $\mathfrak{B}:\mathbb{R}^n \to 2^\mathbb{X}$ is the best response map defined in (\ref{eq:bestresponsemap}).
\end{definition}

\changes{Thus, for any $x^*\in\mathbb{NE}(\mathcal{F}_\mathfrak{P})$ and $p^*=\mathcal{F}_\mathfrak{P}(x^*)$, $x^*$ is a best response to $p^*$ and there is no incentive to deviate from $x^*$, making Nash equilibria a natural notion. See \cite{Romano2024Stackelberg} and \cite{Maheshwari2024Follower-} for recent control-theoretic analyses using alternative notions.}

\begin{remark} \label{rem:NSExplanation} 
Condition (NS) is termed {\bf Nash stationarity} because, for a game  \( p(t)=\mathcal{F}(x(t)) \), it ensures the (EDM) reaches equilibrium precisely when \( x \in \mathbb{NE}(\mathcal{F}) \). Importantly for optimization, if \( \mathcal{F} \) is a potential game with concave potential \( f \), the set \( \mathbb{NE}(\mathcal{F}) \) corresponds to the maxima of \( f \) in \( \mathbb{X} \) (see~\cite[\S3,\S5]{Sandholm2001Potential-games}).
\end{remark}

\section{System-theoretic Passivity Concepts}
\label{sec:MainResult}

In \S\ref{sec:main_results}, we will use two nonstandard passivity notions to establish convergence properties for $\mathbf{x}$. In this section, we will define these notions and explore their basic properties.


\subsection{Counterclockwise dissipativity (CCW)}
We start by adapting to our framework the concept of counterclockwise systems proposed in~\cite{Angeli2006Systems-With-Co} to study stability and robustness of nonlinear feedback systems. 

\begin{definition} The map $\mathfrak{F}:\mathbf{x} \mapsto \mathbf{u}$ is \underline{counterclockwise} dissipative (or {\bf CCW} for short) when
\begin{equation} \label{eq:CCWDEf} \inf_{T>0,\mathbf{x} \in \mathscr{X}} \int_0^T \dot{u}'(t)x(t)dt > -\infty.
\end{equation}
\end{definition} 
 \vspace{.1 in}

\subsubsection{Examples of $\mathfrak{F}$ satisfying CCW}

When $\mathfrak{F}$ is a game $\mathcal{F}$, we know from \cite[Theorem~1]{Martins_2025aa} that $\mathfrak{F}$ is CCW if and only if $\mathcal{F}$ is potential. Modifying the proof of \cite[Theorem~1]{Angeli2006Systems-With-Co}\changes{,} one can show that if $\mathfrak{F}$ is LTI with transfer function $F$ then $\mathfrak{F}$ is CCW if the negative imaginary (NI) condition $j(F(j\omega)-F'(-j\omega))\succeq 0$ holds for all $\omega>0$. The concept of NI for multiple inputs and outputs was proposed in~\cite{Lanzon2008Stability-Robus} to study stability and robustness. Relevant subclasses of NI systems and a dissipative framework for stability are studied in~\cite{Lanzon2023Characterizatio}. 

\begin{remark} \label{rem_conepayoffs} If $\mathfrak{F}$ and $\mathfrak{G}$ are CCW\changes{,} $\xi \mathfrak{F}+\zeta \mathfrak{G}$ is CCW for any non-negative $\xi$ and $\zeta$. \changes{Thus,} the set of CCW maps is a convex cone (see \cite[Proposition~III.1]{Angeli2006Systems-With-Co})\changes{, and} the already large class of (NI) systems~\cite{Petersen2016Negative-imagin} and \changes{CCW} nonlinear generalizations~\cite{Ghallab2018Extending-Negat} can be combined with potential games (see \cite[Theorem~\changes{1}]{Martins_2025aa}) to form a broad convex cone of CCW maps. 
\end{remark}

In the following example, a potential game has an LTI additive perturbation with transfer function matrix $G(s)$. 

\begin{example}\label{ex_NIAndPotential} For a potential game $\mathcal{F}$, a given time-constant $\lambda^{-1}>0$, $b$ in $\mathbb{R}^n$, and $A=A'$ in $\mathbb{R}^{n \times n}$, consider
\begin{subequations}
\label{eq_examplepayoff}
\begin{align}
\dot{q}(t) &= \lambda \big ( Ax(t) + b - q(t) \big ), \qquad t\geq0, \quad q(0)=0, \\
u(t) &= \mathcal{F}\big ( x(t) \big ) + k \lambda \big ( A x(t) + b - q(t) \big)
\end{align}
\end{subequations} where $k$ is a real constant satisfying $k A \preceq 0$.
\end{example}

The transfer function matrix $G(s)$ from $\mathbf{x}$ to $k \lambda ( A \mathbf{x} - \mathbf{q})$ is NI satisfying ${j(G(j\omega)-G'(-j\omega)) = -2 k \lambda^2 \tfrac{\omega}{\omega^2+\lambda^2}A}$, and we conclude from Remark~\ref{rem_conepayoffs} that~(\ref{eq_examplepayoff}) is CCW with stationary game $\mathcal{F}(x)$. Notice that the effect of $b$ in~(\ref{eq_examplepayoff}) vanishes exponentially fast and plays no role in testing for CCW. 

\begin{remark}\label{rem:detailsaboutexampls} Example~\ref{ex_NIAndPotential} generalizes well-motivated examples in~\cite{Fox2013Population-Game}. Namely,\cite[(83)]{Fox2013Population-Game} follows by specializing~(\ref{eq_examplepayoff}) with $k=1$ and $\mathcal{F} ( x )=Ax+b$, where $A \prec 0$. The example \cite[(77)]{Fox2013Population-Game} follows by specializing~(\ref{eq_examplepayoff}) with $k=-\lambda^{-2}$, and $\mathcal{F} ( x )=\lambda^{-1}(Ax+b)$. With this parameter choice, we must impose $A \succeq 0$, such as in a coordination game, in contrast with~\cite[(77)]{Fox2013Population-Game} where $A \prec 0$.
\end{remark}

\subsection{Definition of $\delta$-passivity and $\delta$-antipassivity} \label{subsec:dpass}

Developed to generalize contractive games, \mbox{$\delta$-passivity} is defined below in a form less stringent than, and hence implied by, both the original~\cite{Fox2013Population-Game} and storage-based~\cite{Certorio_2024aa} versions.

\begin{definition} \label{def:dpass}
 We categorize a PDM $\mathfrak{F}$ acting as $\mathfrak{F}:\mathbf{x}\mapsto\mathbf{u}$ as \underline{$\delta$-passive} when for any input $\mathbf{x}$ in $\mathscr{X}$ it holds that
    $$ \inf_{T>0}\int_0^T \dot{u}'(t)\dot{x}(t)dt > - \infty,$$  and $\mathfrak{F}$ is said to be \underline{ $\delta$-antipassive} when $-\mathfrak{F}$ is $\delta$-passive. 
    The EDIM is \underline{$\delta$-passive} when, for any input $\mathbf{p}$ in $\mathscr{P}$ and any resulting solution $\mathbf{x}$ in $\mathscr{X}$ satisfying (EDIM), it holds that
    $$ \inf_{T>0}\int_0^T \dot{x}'(t)\dot{p}(t)dt > -\infty. $$
\end{definition}
\vspace{.05in}
\begin{remark} \label{rem:d-pass_ex}
\cite[Theorem~2]{Certorio_2024aa} guarantees that (EDIM) is $\delta$-passive if either $\beta=0$ or $\mathcal{T}$ can be expressed as $\mathcal{T}=\alpha^\mathrm{SEPT}\mathcal{T}^\mathrm{SEPT}+\alpha^\mathrm{ND}\mathcal{T}^\mathrm{ND}$ for some SEPT rule $\mathcal{T}^\mathrm{SEPT}$, ND-IPC rule $\mathcal{T}^\mathrm{ND}$, and nonnegative weights $\alpha^\mathrm{SEPT}$ and $\alpha^\mathrm{ND}$. Examples of rules leading to a $\delta$-passive (EDIM) include $\mathcal{T}^{\mathrm{\tiny Smith}}$, $\mathcal{T}^{\mathrm{\tiny BNN}}$ and $\mathcal{T}^{\mathrm{\tiny (1)}}$ given in Examples~\ref{example:Smith}, \ref{ex:BNN} and~\ref{ex:hybrid}.
\end{remark}

All contractive games for which $\mathcal{F}$ is continuously differentiable are $\delta$-antipassive \cite{Fox2013Population-Game}. The ``smoothing-anticipatory" PDMs from \cite{Park2019From-Population} provide dynamic examples of $\delta$-antipassive PDMs beyond contractive games.

\begin{remark}
    The set of $\delta$-antipassive maps is a convex cone.
\end{remark}

\section{Main Results}
\label{sec:main_results}


\begin{lemma} \label{lemma} Consider that a $\delta$-passive (EDIM) with well-behaved rule $\mathcal{T}$ and a payoff mechanism $\mathfrak{P}$ satisfying Assumption~\ref{ass:PayoffMech} are given, and that $\mathbf{x} \in \mathscr{X}$ is a solution of (EDIM) in feedback with the payoff mechanism via ${\mathfrak{P}:\mathbf{x} \mapsto \mathbf{p}}$ (Figure \ref{fig:closedloop}). If the following inequality is satisfied
\begin{equation} \label{eq:intbounded}
\int_0^\infty \rho \big( x(t),p(t) \big) dt < \infty,
\end{equation} indicating that the integral is bounded, then it holds that
\begin{equation} \label{eq:dto0}
\lim_{t \rightarrow \infty} d \Big( x(t), \mathfrak{B} \big ( p(t) \big) \Big) = 0,
\end{equation} where $d \big ( x, \mathbb{S} \big) : = \inf_{y \in \mathbb{S}} \| x-y \|$ for $\mathbb{S} \subset \mathbb{X}$. If, in addition to the conditions of the lemma, there is no best response component in (EDIM), that is $\alpha=0$,  then it holds that
\begin{equation} \label{eq:dtoNE}
\lim_{t \rightarrow \infty} d \Big( x(t), \mathbb{NE} \big ( \mathcal{F}_{\mathfrak{P}} \big) \Big) = 0.
\end{equation}
\end{lemma}


\textbf{Proof:} Proposition~\ref{prop:rho} along with Lipschitz continuity and boundedness of $(x(t),p(t))$, where Assumption~\ref{ass:PayoffMech}.2 ensures $p(t)$ is bounded, imply that $\rho(x(t),p(t))$ is Lipschitz continuous. With (\ref{eq:intbounded}), {B}arb\u{a}lat's Lemma~\cite{Farkas2016Variations-on-B} ensures $\lim_{t\to\infty}\rho(x(t),p(t))=0$. Via the limit and continuity of $\rho$, \changes{any} accumulation points \changes{of $(\mathbf{x},\mathbf{p})$} must be zeros of $\rho$. By boundedness of the trajectory $(\mathbf{x},\mathbf{p})$, if (\ref{eq:dto0}) did not hold, there would exist an accumulation point $(x^\star,p^\star)$ of $(\mathbf{x},\mathbf{p})$ such that $d(x^\star,\mathfrak{B}(p^\star))>0$, which would imply $x^\star\notin\mathfrak{B}(p^\star)$, and $\rho(x^\star,p^\star)=0$. However, by Proposition~\ref{prop:rho}, this would present a contradiction as $\rho(x^\star,p^\star)=0$ would imply $x^\star\in\mathfrak{B}(p^\star)$. Thus, we conclude that (\ref{eq:dto0}) must hold.
If $\alpha=0$, the proof proceeds like that of \cite[Theorem 2]{Martins_2025aa}. With continuity of $\mathcal{V}$, (i) (\ref{eq:PC}) and $\lim_{t\to\infty}\rho(x(t),p(t))=0$ imply $\lim_{t\to\infty}\mathcal{V}(x(t),p(t))=0$ and (ii) (\ref{eq:NS}) and $\lim_{t\to\infty}\mathcal{V}(x(t),p(t))=0$ imply (\ref{eq:dtoNE}) holds.~$\square$
\vspace{.02 in}
\begin{theorem} \label{thm:1} \changes{
Let $\mathcal{T}$ be a well-behaved rule for which (EDIM) is $\delta$-passive.   Suppose that the payoff mechanism $\mathfrak{P}$ satisfies Assumption~\ref{ass:PayoffMech}, with $\mathfrak{F}$ being CCW, and at least one of the following holds: (1) $\mathfrak{F}$ is $\delta$-antipassive and $\gamma\nu>0$, (2) $\mathfrak{F}$ is $\delta$-passive and $\gamma\nu<0$, or (3) $\gamma\nu=0$. Then, for any solution $\mathbf{x} \in \mathscr{X}$ of (EDIM) interconnected in feedback with the payoff mechanism $\mathfrak{P}:\mathbf{x} \mapsto \mathbf{p}$ (Figure \ref{fig:closedloop}), it holds that
\begin{equation} \label{eq:TheoremIneq} \int_0^\infty \rho \big( x(t),p(t) \big) dt < \infty, \end{equation} which, using Lemma~\ref{lemma}, ensures (\ref{eq:dto0}). If (EDIM) has no best response component, then~(\ref{eq:TheoremIneq}) also ensures (\ref{eq:dtoNE}).
}
\end{theorem}


\textbf{Proof:} There exist constants $c_1,c_2,c_3\in\mathbb{R}$ such that
\begin{equation}\arraycolsep=1.4pt
\left.
\begin{array}{rl} \label{eq:ineq}
    \int_0^T \dot{p}'(t)\dot{x}(t)dt &\ge c_1 \\[1pt]
    -\mu\gamma(\nu+\mu^{-1})\int_0^T \dot{u}'(t)x(t)dt &\le c_2 \\[1pt]
    \gamma\nu\int_0^T \dot{u}'(t)\dot{x}(t)dt &\le c_3
\end{array} \right\}
\end{equation}
holds for any $T>0$ as (EDIM) is $\delta$-passive, $\mathfrak{F}$ is CCW \changes{and (FOPM) has nonnegative static gain}, and $\mathfrak{P}$ satisfies one of the $\delta$-passivity conditions from the theorem statement.

From Assumption \ref{ass:PayoffMech}.2, integration by parts, and (\ref{eq:ineq}),
\begin{align*}
    c_1 &\le \int_0^T \dot{p}'(t)\dot{x}(t)\changes{dt}= \gamma\int_0^T (\dot{s}(t)+\nu\dot{u}(t))'\dot{x}(t)dt  \\
        &\le \gamma\int_0^T (u(t)-\mu s(t))'\dot{x}(t)dt + c_3 \\
        &= -\mu \int_0^T p'(t)\dot{x}(t)dt + \gamma(1+\mu\nu)\int_0^T u'(t)\dot{x}(t)\changes{dt} + c_3 \\
        &\le -\mu\int_0^T\rho(x(t),p(t))dt + c_2 + 2 \gamma(1+\mu\nu)\beta_\mathfrak{F} + c_3,
\end{align*}
which we rearrange for
\begin{equation*}
    \int_0^T \rho(x(t),p(t))dt \le 2\gamma(\nu+\mu^{-1})\beta_\mathfrak{F} + \frac{1}{\mu}(c_2 + c_3 - c_1),
\end{equation*}
\changes{yielding}
\begin{equation*}
    \int_0^\infty \rho(x(t),p(t))dt \le 2\gamma(\nu+\mu^{-1})\beta_\mathfrak{F} + \frac{1}{\mu}(c_2 + c_3 - c_1),
\end{equation*}
\changes{where the improper integral is guaranteed to exist by the bound and $\rho(x(t),p(t))\ge0$, from Proposition \ref{prop:rho}}. With \changes{the} upper-bound, \changes{(\ref{eq:dto0}) and (\ref{eq:dtoNE})} follow from Lemma~\ref{lemma}.~$\square$

\begin{corollary} \label{cor:1}
    \changes{The results of Theorem \ref{thm:1} hold when $\mathfrak{P}$ is replaced by} $\mathfrak{H}=\changes{\mathfrak{Y}}+\mathfrak{G}$, where \changes{$\mathfrak{Y}$} meets the requirements of Theorem~\ref{thm:1} \changes{imposed on $\mathfrak{P}$ and either} (i) $\mathfrak{G}$ is a CCW, $\delta$-antipassive, \changes{BIBO stable PDM that admits a stationary game}, or (ii) $\mathfrak{G}$ satisfies the conditions \changes{from} Theorem~\ref{thm:1} and \changes{$\mathfrak{Y}$} and $\mathfrak{G}$ each use the same value for $\mu$ \changes{in the associated (FOPM)}.
\end{corollary}

\changes{
\textbf{Proof:} Let $T>0$, $\mathfrak{Y}:\mathbf{x}\mapsto\mathbf{p}_1$, and $\mathfrak{G}:\mathbf{x}\mapsto\mathbf{p}_2$, so $\mathfrak{H}:\mathbf{x}\mapsto\mathbf{p}=\mathbf{p_1}+\mathbf{p_2}$. Begin with the first inequality in (\ref{eq:ineq}) apply $\dot{p}(t)=\dot{p}_1(t)+\dot{p}_2(t)$ to split the integral. For (i), use $\delta$-passivity of $\mathfrak{G}$ to upper bound $\int_0^T\dot{p}_2'(t)\dot{x}(t)dt$ and proceed according to the proof of Theorem~\ref{thm:1} to derive an upper bound on $\int_0^T p_1'(t)\dot{x}(t)dt$. Apply integration by parts and $\mathfrak{G}$ CCW to derive an upper bound on $\int_0^T p_2'(t)\dot{x}(t)dt$. Combine the inequalities for an upper bound on $\int_0^T\rho(x(t),p(t))dt$. For (ii), apply the same steps as in the proof of Theorem~\ref{thm:1} simultaneously to $\int_0^T \dot{p}_1'(t)\dot{x}(t)dt$ and $\int_0^T \dot{p}_2'(t)\dot{x}(t)dt$ and use the shared value of $\mu$ to achieve an upper bound on $\int_0^T\rho(x(t),p(t))dt$. In either case, the proof continues from the upper bound identically to the proof of Theorem~\ref{thm:1}.~$\square$
}

When the conditions of Lemma~\ref{lemma} are met, for which Theorem~\ref{thm:1} and Corollary \ref{cor:1} provide sufficient conditions, (\ref{eq:dto0}) requires that $x(t)$ tends to $\mathfrak{B}(p(t))$. When $\alpha=0$, the result is strengthened to (\ref{eq:dtoNE}). Knowing that $x(t)$ will converge to the Nash equilibrium set of the stationary game can greatly simplify the design of a payoff mechanism as it is only necessary to ensure the placement of the Nash equilibria.

\section{Numerical Example}

To validate our results, we consider a Braess'\changes{s} paradox inspired congestion game with the addition of a dynamic toll, shown in Figure~\ref{fig:pdm_congestion}. We consider \changes{three routes from A to D as strategies}: (1) ABD, (2) ACBD, and (3) ACD. This induces the following payoffs:
\begin{equation*}
    p=\mathcal{F}(x) - \begin{bmatrix}
        0 \\
        b\tau_\mathrm{toll} \\
        0
    \end{bmatrix}, \mathcal{F}(x):=-\begin{bmatrix}
        7 + 2(x_1+x_2) \\
        5 + 5x_2 + 2(x_1+x_3) \\
        7 + 2(x_2+x_3)
    \end{bmatrix}.
\end{equation*}

\changes{Absent} the toll, the Nash equilibrium is $(0.25,0.5,0.25)'$, which \changes{yields a payoff vector of $(-8.5,-8.5,-8.5)'$}. However, the average payoff is improved by implementing a toll to drive the equilibrium \changes{towards} $(0.5,0,0.5)'$, \changes{which would yield} an average payoff of $-8$.

Referring to Corollary~\ref{cor:1}\changes{(i)}, $\mathfrak{G}$ is the congestion game absent the toll, and \changes{$\mathfrak{Y}$} uses $u(t)=-(0,x_2(t),0)'$, $\nu=0$, \changes{and} $\gamma=b$\changes{, using the (FOPM) to impose the toll $\dot{\tau}_\mathrm{toll}(t)=x_2(t)-\mu\tau_\mathrm{toll}(t)$, yielding} the stationary game, ${\mathcal{F}_\mathrm{s}(x) = \mathcal{F}(x) - (0,b\mu^{-1}x_2,0)}$,
with Nash equilibrium $(\tfrac{1}{2}(1+b\mu^{-1}),\ 1,\ \tfrac{1}{2}(1+b\mu^{-1}))'/(2+b\mu^{-1}).$ \changes{Provided $\mu\ll b$, the Nash equilibrium lies near $(0.5,0,0.5)$, so precise knowledge of $b$, a transfer constant mapping tolls to ``equivalent" traversal times, is unnecessary.}

\changes{The simulation \cite{hankins_nash_2025} sets $b=1$, $\mu=100^{-1}$, $\beta=1$ and employs the well-behaved learning rules from Examples \ref{example:Smith}-\ref{ex:hybrid}, which each yield a $\delta$-passive (EDIM) by Remark \ref{rem:d-pass_ex}. The initial conditions are $x^\mathrm{Smith}(0) = (0,0.9,0.1)'$, $x^\mathrm{BNN}(0) = (0,0,1)'$, and $x^\mathrm{(1)}(0) = (0.6, 0.3, 0.1)'$. Each rule-initial condition pair is simulated with $\alpha=0$ and $\alpha=1$. Figure \ref{fig:plot} exhibits the convergence guaranteed by Corollary~\ref{cor:1}.}

\changes{Corollary \ref{cor:1}(i) applies as the (EDIM) is $\delta$-passive and $\mathfrak{G}$ and $\mathbf{x}\mapsto\mathbf{u}$ are each concave potential games, which is true for all congestion games with nondecreasing edge costs \cite[Exercise 3.1.5]{Sandholm2010Population-Game} and can be directly verified for $\mathbf{x}\mapsto\mathbf{u}$, so they are contractive \cite{Hofbauer2007Stable-games}, CCW, and $\delta$-antipassive. Moreover, Assumption \ref{ass:PayoffMech} holds for games as discussed in \S III.}

\begin{figure}
    \center
    \vspace{.02in}
    \scalebox{1}{
\begin{tikzpicture}[scale=1.5]
    \node[circle, draw, fill=yellow!10, thick] (A) at (0,0) {A};
    \node[circle, draw, fill=yellow!10, thick] (B) at (4,0) {D};
    \node[circle, draw, fill=yellow!10, thick] (mid1) at (2,0.6) {B};
    \node[circle, draw, fill=yellow!10, thick] (mid2) at (2,-0.6) {C};

    \draw[->, thick] (A) -- node[above, xshift = -1mm, rotate=16.7] {$5$} (mid1);
    \draw[->,thick] (A) -- node[below, rotate = -16.7] {$2+2(x_2+x_3)$} (mid2);
    \draw[->,thick] (mid2) -- node[right] {$1+x_2+b\tau_\mathrm{toll}$} (mid1);
    \draw[->,thick] (mid1) -- node[above, rotate=-16.7] {$2+2(x_1+x_2)$} (B);
    \draw[->,thick] (mid2) -- node[below, rotate=16.7] {$5$} (B);

\end{tikzpicture}
}
    \caption{Traffic network and latency functions alongside the links.}
    \label{fig:pdm_congestion}
\end{figure}

\begin{figure}
    \centering
    \scalebox{0.95}{
    \input{plot}
    }
    \caption{Plot showing convergence for three distinct learning rule-initial condition pairs. Trajectories with the best response term are dotted.}
    \label{fig:plot}
\end{figure}

\section{Conclusions}

In this paper, we \changes{applied} CCW and $\delta$-passivity to \changes{derive} convergence results for payoff mechanisms with first-order modifications and $\delta$-passive evolutionary dynamics \changes{possibly including} best response dynamics. Prior work employed CCW \cite{Martins_2025aa} and $\delta$-passivity \cite{Park2018Payoff-Dynamic-}, separately, to establish convergence or stability results, but \changes{we have demonstrated} the potential of their joint application\changes{, treating} first-order payoff modifications \changes{more generally than existing work}.

\bibliographystyle{elsarticle-num}
\bibliography{localref} 





\end{document}